# Viewpoints in co-design:

# a field study in concurrent engineering

*Françoise Détienne,* Eiffel Research Group "cognition and cooperation in design", INRIA, Domaine de Voluceau, Rocquencourt, BP 105, 78153 Le Chesnay, France

*Géraldine Martin & Elisabeth Lavigne,* EADS AIRBUS-SA, BTE/SM/GDT-CAO, M0101/9 316 route de Bayonne 31060, Toulouse cedex 03, France

We present a field study aimed at analysing the use of viewpoints in co-design meetings. A viewpoint is a representation characterised by a certain combination of constraints. Three types of viewpoints are distinguished: prescribed viewpoint, discipline-specific viewpoint and integrated viewpoint. The contribution of our work consists in characterising the viewpoints of various stakeholders involved in co-design ("design office" disciplines, and production and maintenance disciplines), the dynamics of viewpoints confrontation and the cooperative modes that enable these different viewpoints to be integrated.

The aim of the study presented in this paper is to analyse the viewpoints brought into play in co-design. The chosen design context is a Concurrent Engineering process. This framework seemed to us to be the most relevant for studying the topic of viewpoint, as the Concurrent Engineering process[1] is assumed to encourage the confrontation of viewpoints during solution development.

Aerospatiale Matra Airbus has conducted the re-engineering of its design processes in a Concurrent Engineering procedure, in order to better control costs, schedules and quality in the design of its products. This industrial development is assisted by cognitive ergonomics research work, which is the framework of this study.

After a state of the art and a presentation of our working hypotheses, we present our field study aimed at understanding the use of viewpoints in an industrial Concurrent Engineering context. Our approach is strongly guided by cognitive ergonomics work on the notion of constraint, and linguistics notions on argumentation.

# 1 State of the art

The confrontation of knowledge and the integration of viewpoints are at the heart of the cooperative mechanisms implemented in co-design. In team design, tasks corresponding to sub-problems are distributed among individuals, each carrying out various sub-tasks. As soon as tasks are divided, conflicts between designers from various disciplines arise and generally negotiation ensues. Design is a process of negotiating among disciplines[2]. Solutions are therefore not only based on purely technical problem-solving criteria. They also result from compromises between designers: solutions are negotiated[3,4].

---

[1] **Darses, F** L'ingénierie concourante : Un modèle en meilleure adéquation avec les processus cognitifs de Conception, in: **P. Bossard, C. Chanchevrier, P. Leclair** (Eds.), *Ingénierie concourante de la technique au social,* Economica, Paris, 1997, pp. 39-55

Viewpoints or views or perspectives have been the focus of research in various disciplines: computer science, linguistics, cognitive ergonomics. An initial general definition of the notion of "viewpoint" is: " for a person, a particular representation of an object". Different participants, with different competencies, skills, responsibilities and interests, inhabit different object-worlds. As such, while admittedly working on the same object of design, they see the object differently[5]. Most of the authors agree that a viewpoint is strongly influenced by the domain area of the designer. Factors such as the field of expertise and specific technical interest play a role in this representation. Several participants see the design object differently according to the constraints specific to their discipline.

---

[2] **Bucciarelli, L L** Engineering design process, in: **F. Dubinskas** (Ed.), *Making time: culture, time and organization in high technology*, Temple University Press, Philadelphia, PA, 1988, pp. 92-122

[3] **Bucciarelli , L L** An ethnographic perspective on engineering design, *Design studies,* 9-3 (1990) 159-168

[4] **Martin, G, Détienne, F and Lavigne, E** Negotiation in collaborative assessment of design solutions: an empirical study on a Concurrent Engineering process, *Proceedings of CE'2000 (International Conference on Concurrent Engineering)* Lyon, France, 2000

[5] **Bucciarelli, L L** Between thought and object in engineering design, *Design Studies,* 23 (2002) 219-231

Our approach is based on several working hypotheses. Firstly we have constructed an operational definition of the viewpoint notion in order to study viewpoints involved in multidisciplinary co-design. We consider that a viewpoint is a representation characterised by a certain combination of constraints: it is strongly influenced by domain area or discipline knowledge. Secondly, we have adopted a dynamic approach to the notion of viewpoint. During the design process, different viewpoints are adopted. In the course of co-design, in particular in multidisciplinary meetings, viewpoints evolve and may become shared by various disciplines designers: integrated viewpoints are constructed.

Furthermore, we have adopted several methodological principles in order to choose our field study situation and to analyse this situation. Firstly, we consider that viewpoints are expressed, more or less explicitly, in multidisciplinary co-design meetings, in particular, during the assessment of solutions. Thus it is on this kind of situation that we have focused our field study. Secondly, owing to the collective nature of the activity, we consider that viewpoints should be expressed, more or less explicitly, through argumentation. Thus it is through the analysis of the argumentation process, its dynamics and the kind of arguments involved that we will focus our analysis of multidisciplinary co-design meetings.

## 1.1 Viewpoints as representations of constraints combination

The most common conception of design problems is to consider them as "ill-structured" problems[6,7]. The specifications given at the start are never complete or without ambiguity: initial problem specifications are not sufficient to define the goal, i.e., the solution, and progressive definition of new constraints is necessary. During the design process, alternative solutions are proposed and the choice of one solution among the set of proposed solution is based on assessment via multiple constraints. The designers develop and assess design solutions partly according to their own specific constraints, which reflect their own specific viewpoints, in relation with the specificity of the tasks they perform and their personal preferences[8,9]. Furthermore the selection and the weighting of constraints evolve through the participants' interactions.

---

[6] **Simon, H** The structure of ill-structured problems, *Artificial Intelligence,* 4. (1973) 181-201

The notion of constraints has been understood from different angles: (1) according to their origin - prescribed constraints, derived constraints; (2) according to their importance – validity constraints and preference constraints[8,10]. The use of a particular

---

[7] **Visser, W** A Tribute to Simon, and some —too late— questions, by a cognitive ergonomist, *Proceedings of the International Conference In Honour of Herbert Simon "The Sciences of Design The Scientific Challenge for the 21st Century"*, Lyon, France, 2002 (edited as INRIA Research Report N° 4462 INRIA, Rocquencourt, Fr)

[8] **Eastman, C M** Cognitive processes and ill-defined problems: a case study from design, In: **D. E. Walker, L. M. Norton** (Eds.), *Proceedings of the First Joint International Conference on Artificial Intelligence*, Mitre, Bedford. MA, 1969

[9] **Chen, A, Diettrich, T G and Ullman, D G** A computer-based design history tool, *Proceedings of the 1991 NSF Design and Manufacturing Conference*, Austin, Texas, 1991

[10] **Bonnardel, N** L'évaluation réflexive dans la dynamique de l'activité du concepteur, in: **J. Perrin** (Ed.), *Pilotage et évaluation des processus de conception,* L'Harmattan, 1999

combination of constraints, characterising a viewpoint, determines the level of abstraction at which the design object is represented[11].

In computer science, authors have characterised viewpoints in ways which are not necessarily related to the representation of constraints. In object-oriented (OO) databases or OO languages, viewpoints act as filters on the representation of an object (see for example[12]). In distributed AI applied in requirement engineering for complex software design, the use of viewpoints is to organise multi-perspective software development and to manage inconsistency (see for example[13]). In AI for design[14], viewpoints are considered as particular combinations of constraints, corresponding implicitly to levels of abstraction. As in this latter approach, we consider a viewpoint characterised by a certain combination of constraints. Adopting a particular combination of constraints should determine the level of abstraction at which the design object is evaluated. In this way a viewpoint would not only act as a filter of the design objects features as in the object-oriented approach but also as a way to switch from one level of abstraction to another[15].

---

[11] **Rasmussen, J** On the structure of knowledge a morphology of mental models in a Man-Machine System context, *Riso-m-2192, Riso national laboratory*, DK-4000 Roskilde, Denmark, 1979

[12] **Abiteboul, S and Bonner, A** Objects and views, *Proceedings ACM SIGMOD, Symposium on the management of data,* 1991

[13] **Finkelstein, A, Kramer, J, Nuseibeh, B, Finkelstein, L and Goedicke, M** Viewpoints: a framework for integrating multiple perspectives in system development, *International Journal of Software Engineering and Knowledge Engineering,* 2-1 (1992) 31-58

[14] **Trousse, B** Viewpoint management for cooperative design, *Proceedings of the IEEE Computational Engineering in Systems Apllication (CESA'98),* Hammamet, Tunisia, 1998

Limitations of previous work are twofold. Firstly, most of the previous works adopt a static approach to the notion of viewpoint. We will adopt a dynamic approach which considers that viewpoints are representations which evolve along the design process. Then a research question is to analyse the dynamic process, in particular the argumentation process, which supports this evolution of viewpoints. Secondly, although most authors refer to the domain knowledge influence on viewpoints, there is no clear analysis of this domain dependency. In our approach of viewpoints as representations characterised by particular combinations of constraints, we will examine the domain specificity of these constraints and also the shared nature of the viewpoint among designers.

## *1.2 Viewpoints expressed through argumentation*

Our working assumption is that viewpoints are expressed through argumentation in multidisciplinary meetings, aimed at co-design, in particular, the assessment of solutions. It is thus on the analysis of these meetings that we have focused our empirical work. Limitations of previous work on collaborative activities involved in co-design meetings concern the analysis of the collective evaluation activities. Despite the great number of previous work on co-design meetings (see for example[16,17,18,19,20]), up

---

[15] **Hoc, J-M** *Cognitive psychology of planning*, Academic Press, London, 1988

[16] **Herbsleb, J D, Klein, H, Olson, G M, Brunner, H, Olson, J S and Harding, J** Object-oriented analysis and design in software project teams, *Human-Computer Interaction*, 10-2 & 3 (1995) 249-292

to now, there is no fine analysis on the assessment modes and the dynamics of evaluation in co-design meetings. The assessment modes of design solutions have been mostly studied in the individual design process. Bonnardel[10] distinguishes between the following three assessment modes in which combination of constraints are adopted in order to evaluate the current solution: (a) analytical assessment mode, i.e., systematic assessment according to constraints; (b) comparative assessment mode, i.e., systematic comparison between alternative proposed solutions and; (c) analogical assessment mode, i.e., transfer of knowledge acquired on a previous solution (accepted or not) in order to assess the current solution.

In collective design, we expect similar assessment modes to be found. Furthermore, due to the collective nature of the assessment process, we expect to observe combined

---

[17] **Walz, D B, Elam, J J, Krasner, H and Curtis, B** A methodology for studying software design teams: an investigation of conflict behaviors in the requirements definition phase, in : **G. M. Olson, S. Sheppard, E. Soloway** (Eds.), *Empirical Studies of Programmers: Second worksho,* Ablex, Norwood, NJ, 1987, pp. 83-99

[18] **Olson, G M, Olson, J S, Carter, M R and Storrosten, M** Small Group Design Meetings: An Analysis of Collaboration, *Human-Computer Interaction,* 7 (1992) 347-374

[19] **D'Astous, P, Détienne, F, Robillard, P N and Visser, W** Quantitative measurements of the influence of participants roles during peer review meetings, *Empirical Software Engineering*, 6 (2001) 143-159

[20] **Stempfle, J and Badke-Schaub, P** Thinking in design teams - an analysis of team communication, *Design Studies*, 23 (2002) 473-496

assessment modes: in this case, each participant in an assessment co-design meeting may use one or several assessment modes in order to convince the other participants.

With respect to linguistic work on argumentation[21,22,23] we will consider that these modes involve, in a meeting situation, the use of various types of arguments. Linguists distinguish different kinds of arguments: argument by comparison, argument by analogy, argument of authority. An argument by comparison compares several objects in order to assess them in relation to each other. Arguments by analogy are arguments that highlight a precedent, i.e., they enable the present case to be compared to a typical case proposed as a model. We consider that the comparative assessment mode and the analogical assessment mode may involve what linguists call argument by comparison or argument by analogy. Most of these arguments can take the status of argument of authority depending on factors which give a particularly strong weight to the argument. Argument of authority is an indisputable argument that is built on a quotation of statements, so it is in no way a proof, even if it is presented as such. In general, the proposer's argument is the fact that it has been expressed by a particular authorised person, on whom he/she relies, or behind whom he/she hides.

---

[21] **Plantin, C** *L'argumentation,* Seuil, 1996

[22] **Perelman, C and Olbrechts-Tyteca, L** *Traité de l'argumentation,* Ed de l'université de Bruxelles, 1992

[23] **Baker, M J** The function of argumentation dialogue in cooperative problem-solving, in: **F. H. van Eemeren, R. Grootendorst, J. A. Blair, C. A. Willard** (Eds.), *Proceedings of the 4th International Conference on Argumentation (ISSA'98)*, SIC SAT Publications, Amsterdam, 1998, pp. 27-33

Limitations of the linguistic and pragmatic approach concern the kind of situations studied. It has not been applied to knowledge-rich situations such as collective design in which the various disciplines of the participants should strongly influence the argumentative process and the viewpoints involved in this process.

## 1.3 *Negotiation patterns and viewpoints integration*

In the argumentative dialogue, a proposer will express a proposal that will be argued about by presenting a certain amount of information substantiating the initial proposal. When everyone has a joint will to reach agreement, we shall talk about negotiation. Negotiation does not force a person to accept a solution, dialogue makes it possible to go towards one conclusion rather than another. For example, the conclusion can be a compromise between what each person wants.

Directly related to our team design situation, the design rationale (DR) approach in computer science have tempted to make explicit the reasoning behind design[24,25,26]. Several DR notations have been developed to express the design reasoning as "arguments" about "issues". Among them, QOC and IBIS are probably the most well-known. The QOC notation[27] distinguishes between Questions, Options and Criteria. The Design Space Analysis (DSA) approach that uses QOC consists in creating an explicit representation of a structured space of design alternatives and the considerations for choosing among them. It is a process of identifying key problems (Questions), and raising and justifying (via Criteria) design alternatives (Options).

---

[24] **Buckingham, Shum S and Hammond, N** Argumentation-based design rationale: what use at what cost? *International Journal of Human-Computer Studies,* 40 (1994) 603-652

[25] **Concklin, E J and Burgess, K C** A Process-Oriented Approach to Design Rationale, *Human-computer Interaction,* 6 (1991) 357-391

[26] **Moran, T P and Carroll, J M** *Design rationale: concepts, techniques and uses* Erlbaum, Mahwah, NJ (1996)

A limitation of this approach concerns the lack of viewpoints representation. Whereas arguments refer to criteria or constraints and have some weighting mechanism, combining constraints to make up viewpoints explicit is not possible. The same limitation is also present in cooperative systems such as argumentative systems (for example, see[28] also based on this approach).

Furthermore the generic model of negotiation implemented in these systems does not identify negotiation patterns with respect to particular team situation such as co-design. With our dynamic approach to viewpoints, we will examine the negotiation patterns leading the participants to converge and the modes of viewpoint integration.

## 1.4 Research questions

The contribution of our work consists in characterising the viewpoints of various stakeholders involved in co-design (in our study: "design office" designers, and production and maintenance disciplines), the dynamics of viewpoints confrontation and the cooperative modes that enable the construction of integrated viewpoints.

---

[27] **MacLean, A, Young, R M, Bellotti, V and Moran, T** Questions, Options, and Criteria: elements of design space analysis, *Human-Computer Interaction*, 6-3&4 (1991) 201-250

Considering viewpoints as representations of constraint combinations, our questions are: what are the mechanisms of viewpoints clarification? are there particular constraint combinations which reflect particular status of viewpoint with respect to their shared collective nature?

Considering that viewpoints are brought up through assessment modes, our question is: which assessment modes?

Considering the dynamics of viewpoint confrontation and integration, our questions are: is there a temporal organisation of viewpoints confrontation? what are the integration mechanisms?

# 2 Methodology

## 2.1 Context

We conducted this study during the definition phase of an aeronautical design project, lasting three years, in which the participants work in Concurrent Engineering to design the centre section of an aircraft. These participants use Computer Aided Design (CAD) tools and a Product Data Management System (PDM). About 400 people with ten different disciplines are involved. These disciplines are the traditional design office disciplines (structure, system installation, stress), disciplines that used to intervene further downstream (maintainability, production) and new disciplines that have appeared with the introduction of CAD and PDM tools.

---

[28] **Lonchamp, J** A Generic computer Support for Concurrent Design, *Proceedings of Advances in Concurrent Engineering, CE'2000* Lyon, Fr, 2000

## 2.2 First phase

### 2.2.1 Collection of data during meetings

All the disciplines work on the same part of the aircraft but each person according to his technical competence. "Informal" multidisciplinary meetings are organised, as needed, to assess the integration of the solutions of each discipline into a global solution. We took part in seven of these meetings as observers:

- Four meetings involved upstream design office designers, i.e., designers from structure and designers from systems installation (SI); structure/electricity-SI meeting; structure/hydraulics-SI meeting; structure/flight-control-SI meeting; structure/fuel-SI meeting.

- Two meetings involved upstream-design office designers (structure) and downstream designers: structure/production meeting; structure/maintenance meeting.

On the basis of audio recordings and notes taken during the meeting, we transcribed the full content of the meetings. Each meeting involved three to six designers.

### 2.2.2 Coding scheme

The protocols resulting from the transcriptions were broken down according to the change of locutors. Each individual participant statement corresponds to a "turn". Each turn was coded according to the following coding scheme and broken down again as required to code finer units. Our coding scheme comprises two levels, a functional level and an argumentative level.

The functional level highlights the way in which collective design is performed. Each unit is coded by a mode (request/assertion), an action (e.g., assess) and, an object (e.g., solution n). At this level, a turn can be broken down into finer units according to whether there is a change in mode, activity or object. This level of analysis will not be developed further as our focus in this paper is on the argumentative level. The reader interested in having more detail on the functional level can refer to[29].

The argumentative level brings out the structure of discourse on the basis of a dialogue situation. We coded the proposals for solutions and the different types of arguments used by the designers. Functional units (but not all as some units clearly do not belong to this process) were assigned four kinds of role in the argumentative process:

- Proposal X: solution X is proposed by one or several participants;

- Question: Questions about the proposal are made;

- Argument +(X) or – (X): arguments supporting or not supporting the proposal are advanced by the participants;

- Resolution: the proposal is accepted or rejected by all the participants or there is an absence of conclusion.

We detected converging moves (agreement between participants on the acceptance or rejection of a proposal) and diverging moves (several examples will be given in 3.1). The nature of the arguments was further refined. In particular we examined whether:

- one or several constraints were used in the argument;

---

[29] **Darses, F, Détienne, F, Falzon, P and Visser, W** *A method for analysing collective*

- an example was brought out to convince the others (argument by example used in analogical evaluation);

- the argument had the status of argument of authority. In this case, an argument is presented as inconstestable and therefore it has a particularly strong weight in the negotiation process. An argument can take the status of argument of authority depending on : the status, recognised in the organisation, of the discipline that expresses it; the expertise of the proposer; the "shared" nature of the knowledge to which it refers.

We also constructed design rationale graphs in order to count the number of alternative solutions (options) evoked by the participants for each problem (questions), and the arguments given in favor or against each solution (criteria).

## 2.3 Second phase

### 2.3.1 Auto-confrontations with coded protocols of meetings

We conducted interviews afterwards with the various participants of meetings to validate the coding we had made and make explicit a certain amount of information that was implicit in the meetings. As our focus was on the analysis of viewpoints through the arguments expressed during evaluation meetings, in particular through the notion of constraints, our primary concern was to validate our coding of the argumentation process.

We gave to each participant our coding of the meeting(s) where he/she took part, and asked him/her to assess our coding and to make explicit the case where one or several constraints were implicit in an expressed argument but in fact founded the argument

*design process,* Research Report n°4258, INRIA-Rocquencourt, Fr, 2001

itself. This allowed us to make appear, in the argumentation process, the distinction between:

- Argument with explicit constraint(s): e.g *"If we have a 160mm pulley, and we've only got 140, were going to have a problem "* (explicit system-installation constraint)

- Argument with implicit constraint(s) : e.g *"this fractured on the other aircraft"* (the implicit constraints are stress and structure)

## *2.3.2 Tests with constraints*

Our second concern was to identify what representation each discipline had about constraints: in particular the representation of the meaning assigned to a constraint expressed a certain way and the ordering between constraints. Based on previous work, we thought that these representations may depend on the expertise of the designers, in particular their discipline, but also on the context (the problem-situation addressed). Thus, our tests were constructed depending on the problem-situation, i.e. the meeting, in which constraints had been used.

For each meeting, we collected the constraints used (either explicitly or implicitly) and presented the list to each participant of this meeting. Our question concerned:

- for each constraint: to give their meaning;

- for all constraints: to order them as a function of their importance in this design-problem-situation.

# 3 Results

The six meetings have been segmented into ten design-assessment-sequences and five coordination-sequences. A design-assessment-sequence corresponds to the discussion related to a particular design sub-problem and the proposed solutions corresponding to that problem. One coordination-sequence corresponds to the discussion related to coordination and management of the project (distribution and allocation of tasks). The coordination-sequences will not be considered in the following analysis.

For each design-assessment-sequence analysed, whatever the problem involved, a solution is proposed by a discipline D1. This solution is called the initial solution. D1-designers give arguments to support it in order to convince the other discipline, D2. This solution may be accepted immediately by D2-designers who is convinced of the pertinence of the solution. On the other hand, D2-designers could refuse it, which is the most frequent case. Then follows a negotiation between the two disciplines in order to reach an agreement. An alternative solution is then proposed by D1-designers or D2-designers which will in turn be assessed. Often, several alternative solutions are proposed before a negotiated solution is reached. In our ten design-assessment-sequences, the number of alternative solutions proposed varied between 0 and 21 (as shown later in Table 4). Finally, it sometimes happens that the meeting does not enable a result to be achieved. Each discipline must then work again before another meeting is convened.

## 3.1 Characterisation of viewpoints

Viewpoints have been analysed through the argumentation process. We analysed in which conditions constraints composing viewpoints were made explicit in this process.

Furthermore, we analysed the meaning and weighting of constraints constructed by each discipline.

## 3.1.1 Mechanisms of viewpoints clarification

Constraints can be explicit or implicit in the viewpoint as it is expressed by a designers' argument. The mechanism of viewpoint clarification or explicitation may depend on several dialogue factors. Of course, it may depend on an explanation request made by another participant of the meeting: this is a rather straightforward mechanism that we observed in a systematic way. It may also depend on the speaker assuming shared knowledge with other participants: this factor was difficult to assess. We have identified two other conditions of constraint clarification: diverging move between disciplines; reinforcement of intra-discipline-consensus.

Table 1 gives an example of an implicit divergence which leads a participant to make explicit constraints he used in previous arguments. This table highlights the chronological sequence of the situation at a structure/system installation meeting. The Structure-designer (St1) put forward arguments for rejecting a solution which was proposed previously by a system-installation-designer (SI1). St1 does so by referring to a similar problem, saying: "*This fractured on the other aircraft* [Arg 14 Stress/structure constraints]". Even if this argument is founded on two constraints, stress and structure, these constraints remain implicit in what is said by the designer. Faced with the lack of reaction from the SI-designer, St1 argues still further: "*Why? Because according to the computation there was a relative displacement of the beam of approximately 2mm with respect to the other one* [Arg 15 Stress/structure constraints]". This is reformulated by St2 [Arg16].

The lack of reaction from the SI-designer led the Structure-designer to assume an implicit divergence on the part of the SI-designer. He thus felt compelled to argue his rejection of the solution by making his viewpoint explicit (the stress and structure constraints). This divergence is, moreover, explained just after the 83rd successive contribution to the discussion, with the following words of SI1: "*So we contacted several people dealing with the electrical installation on the other aircraft, and had no feedback of any incidents at that level*".

Table 2 shows, in chronological order a strengthening of the consensus of opinion by another representative from the same discipline. In this structure/system-installation meeting, the two Structure-Designers reject a solution proposed by the SI-Designers. To show his disagreement with the SI-designers, St(1) puts forward his arguments for rejecting the solution (78th contribution to the discussion): "*Because in that case they would have to do another study* [Arg 3 constraints relating to program and study deadlines and costs] *and add material* [Arg 4 stress and structure constraint]". St(2) goes even further than St(1) by explaining the design constraint (program-study constraint) left implicit by St(1). He says: "*The complete study already conducted will have to be done again* [Arg 5 constraints relating to program and study deadlines and costs], *as there is an offset of the box beam* [Arg 6 stress-constraint and structure-constraint]".

By using this process for strengthening the consensus of opinion, St(2) backs up what St(1) has already said. He emphasises this mechanism by using two arguments (arg5 and arg6) that refer to the same constraints used by St(1) in arguments 3 and 4. By doing this he obliges the SI-Designers to justify the advantages of the solution they propose even further. By using these means, the Structure-designers endeavour to impose their viewpoint.

## 3.1.2 Meaning and weighting of constraints

Constraints used in the argumentation process are of two kinds: prescribed constraints independent of the discipline: those constraints are prescribed in the design specification and, *a priori*, shared by all the players of the design process; derived constraints specific to a discipline. We found that, even though some constraints used by different players in a meeting are the same at a surface level (same terminology), these constraints may have different meanings in the viewpoints expressed by players from different disciplines. Also, the level of refinement selected may be different according to the discipline.

## 3.1.2.1 Selection of a meaning for a discipline-independent constraint

We observed that the same constraint (the same terms are used by different players in a meeting) can have different meanings according to the speaker's discipline.

In this case it is necessary to distinguish the two slopes of the sign, the signifier and the meaning. The meaning can have the same generic seme for different speakers but a very different functional seme. Figure 1 illustrates that a cost constraint can, for one discipline, mean "production cost" and, for another discipline, mean "design cost". It seems particularly true for general constraints prescribed for all the players of the design process (e.g., the cost) as opposed to constraints derived by a discipline (e.g., structure).

## 3.1.2.2 Selection of a refinement level in a hierarchical network of a discipline-dependent-constraint

Some constraints expressed in the argumentation process may be organised hierarchically along different levels of refinement. For example, a maintenance constraint may be refined as three constraints: accessibility constraint, dismounting constraint and mounting constraint. However, when we analysed the discipline-dependent constraints used for expressing the viewpoints of different players, we identified some gaps between the level of refinement selected and used in the argumentation process according to the speaker's discipline. For a constraint specific to a discipline, the level of refinement adopted by this discipline is more detailed than the one adopted by the other discipline. Two examples are given in Figure 2.

## 3.1.2.3 Constraints weighting

On the basis of our tests with constraints (second phase) we showed that the constraints weighting, is affected by two factors: the participant's discipline and the design-problem situation. In general, constraints taken into account in a particular meeting are those constraints specific to the disciplines involved in the meeting in addition to the prescribed constraints. However discipline-dependent-constraint weighting depends on speaker discipline. Whereas we found a high intra-discipline agreement on constraint weighting, we found disagreement between disciplines. An example is given in Table3.

The constraints which are specific to hydraulic-system-installation-designers are : system installation and frontier. The constraints which are specific to structure-designers are: structure and stress. We can see that, even if most of these constraints are used by the two disciplines involved in the meeting, the way each discipline orders those constraints by importance is different. Each discipline ranks his/her own constraints as more important than the constraints of his/her interlocutors. Furthermore, we can see in this example, that some constraints are used only by one discipline: time, growth of problem, frontier are used only by hydraulics-designers.

Of course, constraints weighting also depends on the problem in hand. For example, we observed for the same discipline, air system installation, that constraint weighting varied between two problems processed sequentially in a meeting: the maintainability constraint was ranked 3 for problem A and 1 for problem B. Furthermore the production constraint was evoked only for problem A.

## 3.2 Dynamics of viewpoint confrontation

Dynamics of viewpoint confrontation has been analysed with respect to assessment modes. We have found the existence of analytical, comparative or analogical assessment modes in these meetings. This type of result is similar to the assessment modes analysed in individual design[10]. In addition, we have highlighted combined assessment modes, e.g. analytical/analogical. Table 4 shows the occurrence of these assessment modes in the ten design-assessment sequences of the six meetings we have analysed. It shows also that in eight sequences, participants converged on a choice of solution.

Another research question was whether there is a typical temporal organisation of these assessment modes. In 90% of the observed sequences, we found that the different assessment modes are used in the following order:

- Step1: analytical assessment mode of the current solution;

- Step 2: if step 1 has not led to a consensus, comparative or/and analogical assessment is involved;

- Step 3: if step 2 has not led to a consensus, one (or several) argument(s) of authority is (are) used.

We present the combined assessment modes and illustrated them graphically through examples.

## 3.2.1 Analogical /analytical assessment

This mode combines analogical assessment and analytical assessment. In the framework of analogical reasoning, the current solution (the one which is proposed for evaluation) is called the target solution whereas the analogical solution (a previous solution which is brought up in the argumentation process) is called the source solution. Figure 3 illustrates graphically such a combined assessment mode. In this example, D1-designers (designers from one discipline) use the analogical/analytical assessment to convince D2-designers (designers from another discipline) to accept the solution S1 proposed by D1.

D1 designers propose a solution, the target solution S1, which is rejected by D2-designers. In order to convince D2-designers of the adequateness of S1, D1-designers make reference to an analogical solution, the source solution S2. S2 is a solution which was accepted in a past context. In this context S2 was a solution negotiated between D1-designers and designers from another discipline, D3-designers : even if this solution was not so easy to use by D3-designers (this solution was not ideal in terms of some constraints important for these disciplines), they finally accepted it. In their argumentation, D1-designers analyse the source solution S2 according to a set of constraints (analytical assessment). They make explicit positive arguments as well as negative arguments and defend the idea that the D3-designers were able, in the past, to accept this evaluation and therefore the source solution S2. The conclusion of this negotiation process is the acceptance, by D2, of the target solution S1.

## *3*.2.2 Comparative/analytical assessment

This mode combines comparative assessment and analytical assessment. The comparative assessment mode involves systematic comparison between the current solution and one or several alternative proposed solutions. These solutions are alternative to the current proposed solution (the one originally to be assessed). Figure 4 illustrates graphically such a combined assessment mode. In this example, each discipline will propose his own alternative solution. None of them accept the current proposed solution.

D1-designers propose an alternative solution Salt 1 whereas D2-designers propose another alternative solution : Salt 2. Each alternative solution is then analytically analysed by participants of both disciplines.D1-designers positively assess Salt1 (their own proposed alternative solution) and negatively assess Salt2. Conversely, D2-designers positively assess Salt2 (their own proposed alternative solution) and negatively assess Salt1. These analytical assessments allow each discipline to compare the suitability of the two alternative solutions according to various design constraints. In doing so, each discipline makes explicit the design constraints which are judged more important in his field. The conclusion of this negotiation process is that neither of the two proposed alternative solutions are accepted. Rather, a third alternative solution, which is a compromise between Salt1 and Salt2, is generated.

## 3.2.3 Comparative/analogical assessment

This mode combines comparative assessment and analogical assessment. Figure 5 illustrates graphically such a combined assessment mode. In this example, disciplines will propose an alternative solution (comparative assessment) and will defend this solution in reference to a previous source solution which was accepted in the past (analogical assessment).

D1-designers propose and defend the current solution S1. D2-designers propose an alternative solution Salt1. In order to defend this alternative solution, they make reference to a source solution, accepted in a past context, which is analogical to Salt1. This source solution is then analogically assessed by the different disciplines. This evaluation allows the disciplines to compare the advantages (positive arguments) and drawbacks (negative arguments) of the current solution S1 and its alternative solution Salt1. D1-designers give negative arguments toward Salt1 based on negative arguments toward the source solution ; this allows them to show, by comparison, the advantages of solution S1. Conversely, D2-designers give positive arguments toward Salt1 based on positive arguments toward the source solution ; this allows them to show, by comparison, the drawbacks of solution S1. The conclusion of this negotiation process is the absence of any negotiated solution or any consensus. In fact, due to the disagreement between the disciplines on the source, a task is planned in order to verify information related to the source solution. The design rationale about the source solution has to be reconstructed for the next meeting.

## 3.3 Cooperative modes for integration

We found that three types of viewpoints were introduced in an invariant order in the argumentation process. Furthermore, we found three integration mechanisms which leaded the participant to converge toward an agreement.

## 3.3.1 Types of viewpoints

A viewpoint is a representation of a combination of constraints. These representations are of three types which have different status according to their shared nature by the team and the prescribed versus constructed nature of the constraints:

- "Prescribed viewpoints" are representations of prescribed constraints, strongly based on the design specification. They are shared by the whole team whatever the discipline of its members. However the meaning and weighting of constraints constructed by various discipline designers may vary.

- "Discipline-specific viewpoints" are representations which are constructed according to discipline specific knowledge (constraints specific to a discipline) and they are shared by designers of the same discipline.

- "Integrated-viewpoints" are representations which are constructed through the argumentation process of the team and they are shared by designers whatever their discipline.

Representations of combinations of prescribed-constraints are referred to as prescribed viewpoints. They are constructed on the basis of the design specifications by all the designers. However, we have seen in section 3.1.2 that these apparently shared representations may hide some gap between the meanings that each discipline associate with these constraints.

Discipline-specific viewpoint are of two kinds: either combination of own discipline-dependant-constraints or combination of own discipline-dependant-constraints with discipline-independent-constraints. Designers use combination of their own discipline-dependant-constraints in their viewpoint. For example structure-designers may use combination of constraints such as structure and stress.

Two variants of combination of own discipline-dependant-constraints with discipline-independent-constraints were observed:

- During the argumentation process: Designers of discipline-1 may use this kind of combination with discipline1-dependant-constraints they consider less important: it is a "weak" discipline-specific viewpoint. We argue that this combination is a way to have one's own point of view accepted by the interlocutors. Indeed, associating one's own constraints, in particular those weighted as less important, with prescribed constraints accepted (and not contestable) by the various disciplines is a way to make a stronger argument in the argumentation process.

- At the end of the argumentation process: Designers of discipline-1 use this kind of combination with discipline1-dependant-constraints they consider more important: it is a "strong" discipline-specific viewpoint. It is a way to check that constraints are satisfied as a result of the negotiation, in particular the prescribed ones and one's own constraint weighted as more important.

Combinations of discipline1-dependant-constraints and discipline2-dependant-constraints represent integrated viewpoints. However we have seen in 3.1.2 that constraints may be weighted differently by players of discipline-1 and players of discipline-2. For example, we observed that structure-designers and system-installation-designers may construct an integrated viewpoint composed of a structure constraint and a system-installation constraint. However, their weighting of these constraints is quite different: the structure-constraint is the highest for the structure-designers whereas the system-installation-constraint is the highest for the system-installation-designers.

## *3*.3.2 *Dynamics and integration mechanisms*

Figure 6 shows the dynamics of viewpoints confrontation/integration. In a first step, each discipline evaluates the proposed solution with an analytical assessment mode. This is done on the basis of a combination of prescribed constraints or a combination of one's own discipline-specific constraints. In this way, prescribed viewpoints or discipline-specific viewpoints are expressed. If this process does not allow the various players to converge, which is generally the case, then a second step occurs.

In the second step, each discipline evaluates the proposed solution with an analytical/comparative assessment mode or analytical/analogical assessment mode. This is done on the basis of a combination of discipline-independent constraints and one's own discipline-specific constraints. In this way, a discipline-specific viewpoint is presented. Or this is done on the basis of a combination of discipline1-specific constraints and discipline2-specific constraints. In this way, an integrated viewpoint is constructed.

We found that three mechanisms were involved in sequences where converging among participants occurred and so viewpoints were integrated.

- Mechanism 1: it consists in explicitly negotiating constraints when the search for alternative solutions is in an impasse;

- Mechanism 2: it consists in evoking shared knowledge concerning the evaluation of a source (previous solution developed for an analogous problem) or a previous alternative solution for which an integrated viewpoint was found;

- Mechanism 3: it relies on argument of authority.

The first mechanism of viewpoint integration consists in constraints negotiation. It has been observed in two design-assessment sequences (sequences 8 and 10) where many alternatives were produced (respectively 15 and 21) and led to an impasse. Alternative solutions were generated and refined and none satisfied the discipline-specific viewpoints and the prescribed viewpoint.

For example, in a problem-solving sequence of the structure-maintenance meeting (sequence 10), six alternatives were first produced then refined leading to a total of twenty-one alternative solutions evokes. The designers recognised that they were not able to produce any more alternative solutions and one of them said « *it cannot work!* ». In order to go out of this impasse, we observed that the designers started negotiation about the constraints themselves. Each discipline states its most important constraints. The structure designers say: "*It must be sealed* [sealing constraint] *and the pitch must be tight* [production constraint]". The system installation designers say: "*It must be installable and removable* [maintainability constraint]". Then each of the two disciplines will make concessions on one of his constraints. Structure designers, who wanted a tight pitch, accepts that it is less so: "*Yes, we can skip one* [pitch] *provided we know it*". Similarly, Maintenance designers stress that the removability constraint is less of a problem "*We take it because it happens once every two times in the life of an aircraft*". After this negotiation, the disciplines get out of the impasse and two alternative solutions among the six generated first are considered acceptable. An additional study will have to be made in order to choose one of the two alternative solutions.

The second mechanism of viewpoint integration occurred in five design-assessment sequences (4, 5, 7, 9, 10) and was the most frequent mechanism leading to convergence. It consists in evoking shared knowledge concerning the evaluation of a source (previous solution developed for an analogous problem) or a previous alternative solution for which an integrated viewpoint was found. This is typically the case in analogical assessment, when the participants in the meeting have shared knowledge about the source solution, e.g., everybody agrees that it works in this similar context. Arguments by analogy served the analogical assessment of the current solution. In this case, there is a transfer of the result of the assessment of an analogical solution (source) developed in the past for the same design project or for a previous design project to the current proposal (target). Whenever the participants shared knowledge about this past design, the argument by analogy was likely to have the strongest weight in the argumentation as illustrated in 3.2.1 (Figure 3).

In this case, the shared knowledge about the past design consists in: the attributes of the source solution; the results of its evaluation process; but, most importantly, the various discipline-dependent constraints used to assess it, the combination of these constraints as it was negotiated in the past design: it is the "integrated viewpoint" reached by the team in the past.

Evoking attributes of the source and the results of its evaluation is quite classical in analogical reasoning: the distinction here is that it is based on knowledge shared by the team of designers, what is called the common operative referential[30] or shared context[31]. In some cases, we observed that participants do not share knowledge about the source which was the case in sequence 3 (example given in 3.2.3 and Figure 3). When knowledge about the past design is not shared, either traces of the past design process are sought, which takes generally much time (in the example given in 3.2.3 it took three months) or an argument of authority (relying on the expertise or the status of the person who enunciates it) is involved.

Argument of authority are used and allow a discipline to "impose" one's own viewpoint. It occurred in two design-assessment sequences (1 and 2). We have found that an argument can take the status of argument of authority depending on :

- the status, recognised in the organisation, of the discipline that expresses it: it was the case for the structure-discipline. In the traditional design process, before concurrent engineering was set up, the structure-designers intervened before system-installation designers and thus their solutions provided the specifications for system-installation designers work. In the new context, we observed that the structure-designers had a high status recognised by the system-installation designers.

---

[30] **De Terssac, G and Chabaud, C** Référentiel opératif commun et fiabilité, in: **J. Leplat, G. de Terssac** (Eds.), *Les facteurs humains de la fiabilité dans les systèmes complexes*, Octarès, Paris, 1990

[31] **Hutchins, E and Klausen, T** Distributed cognition in airline cockpit, in : **Y, Engeström. D, Middelton** (Eds.), *Cognition and communication at work*, Cambridge University Press, Cambridge, 1996

- the expertise of the proposer. The argument is going to make reference to a person recognised by all to be an expert in the discipline. It will be something like " *It's Alphonse who said it would be more logical like that to pick up on these parts of the stringers*".

Several remarks may be made on the basis of these results. Firstly, mechanisms 3 (argument of authority) was an effective mechanism leading to integration of viewpoints only in meetings involving design office designers (sequences 1 and 2). This refers to the "special" status of the structure designers with respect to the system-installation designers. Secondly, mechanism 2 (evoking shared knowledge about past design) was the most effective mechanism leading to integration of viewpoints whatever the disciplines composing the meeting. However, we observed for sequence 10 that this mechanism was not sufficient in the context of a meeting composed of design office designers and maintenance designers. In this case, mechanism 1 (negotiation of constraints) was also involved to lead designers to converge on a solution.

## 4 Discussion

This paper has presented an attempt to analyse viewpoints involved in design within an ergonomic theoretical framework. Viewpoints have been analysed through the argumentation process. The originality of our work is to characterise the viewpoints of the various designers involved in co-design, the dynamics of viewpoints confrontation and the cooperative modes that enable these different viewpoints to be integrated.

We have adopted an approach considering viewpoints as representations which may be shared by several participants. This approach has enabled us to distinguish various types of viewpoints (prescribed, discipline-specific and integrated) depending on their shared nature by the team and the prescribed versus derived nature of the constraints founding them. This approach is also important to identify the evolution of viewpoints in the co-design meetings, from prescribed viewpoints to integrated viewpoints whenever the participants converge toward an agreement.

We have also adopted a dynamic approach on viewpoint. We have shown that the evolution of viewpoints is supported by a typical temporal assessment pattern composed of three steps: (1) analytical assessment; (2) comparative and/or analogical combined assessments: (2) argument of authority.  Three mechanisms of integration have been identified: negotiation of constraints; evocation of shared knowledge on a source solution or a previous alternative solution; argument of authority.

Limitations of our work concern the particularity of the meetings and the field area concerned in our field study. Further studies should be made to verify the generality of our results. Furthermore, the analysis of the argumentative dialogues could be improved by a finer methodology based on argumentative indicators in the language.

However our results can be a basis to specify co-design meetings methodology and support for meetings such as argumentative system. For example, we believe that it is important to support in some way the distinct assessment modes. Furthermore, methodology should encourage viewpoints explicitation: combination and weighting of constraints, with a specific reference to the disciplinary aspects, should be made explicit both for supporting the argumentative process and for ensuring the traceability of design decisions.

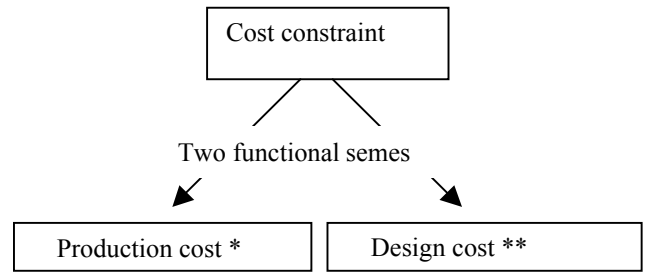

*Figure 1  Selection of a meaning for a discipline-independent-constraint*
KEY
* MEANING ACCORDING DESIGN OFFICE FIELD 1
** MEANING ACCORDING DESIGN OFFICE FIELD 2

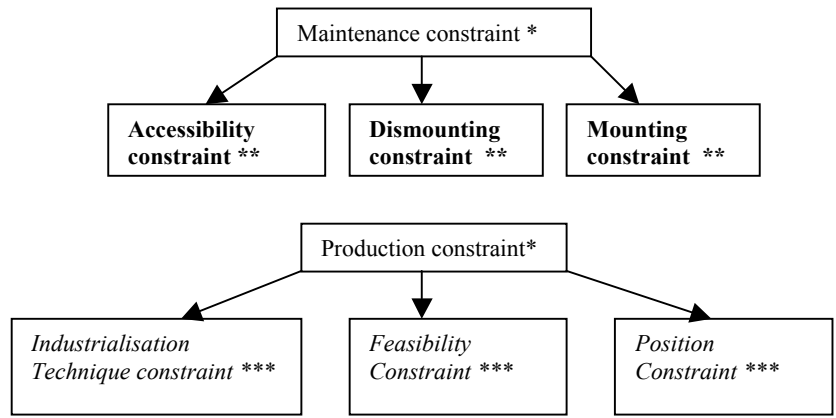

*Figure 2  Selection of a level of detail in the hierarchical network of constraints specific to a discipline*
*KEY*
* CONSTRAINTS EXPRESSED BY DESIGN OFFICE FIELD
** **CONSTRAINTS EXPRESSED BY MAINTENANCE**
*** *Constraints expressed by production*

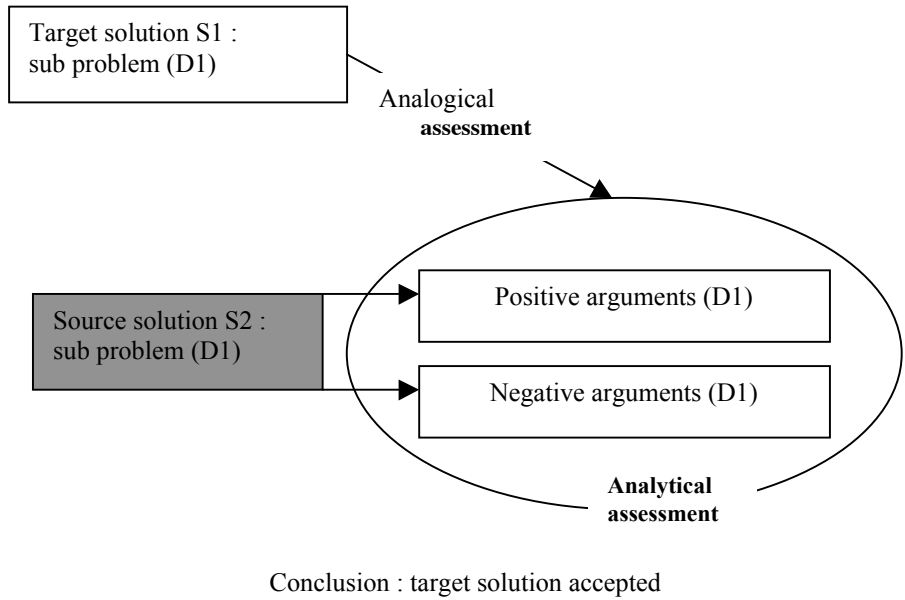

*Figure 3  Analogical /analytical assessment*

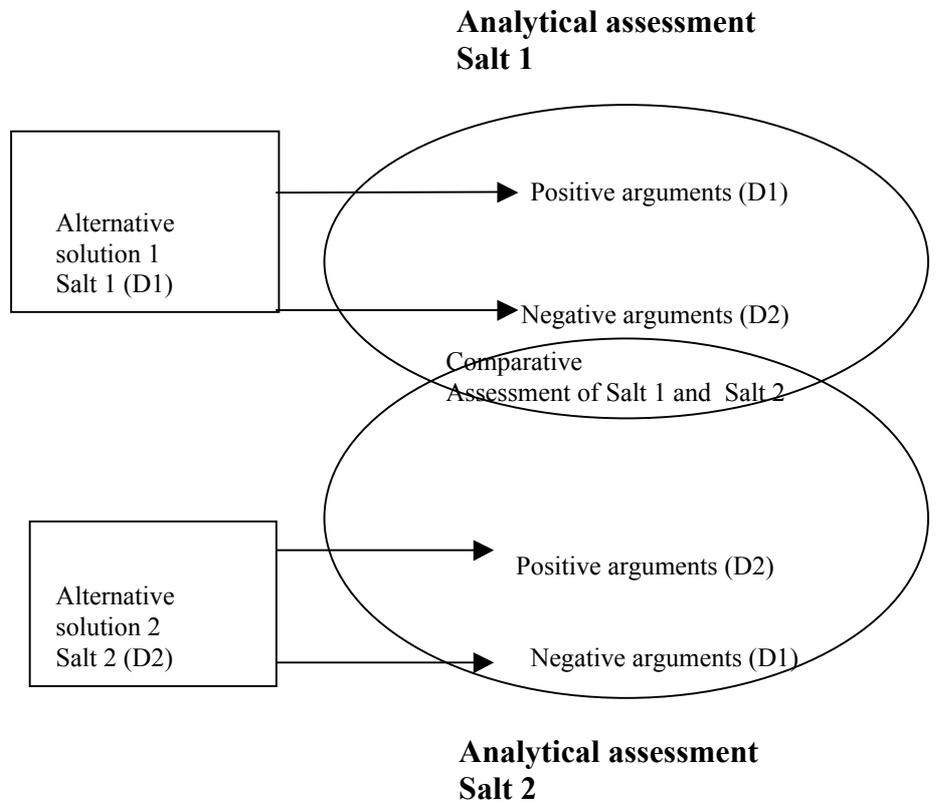

*Figure 4  Comparative/analytical assessment*

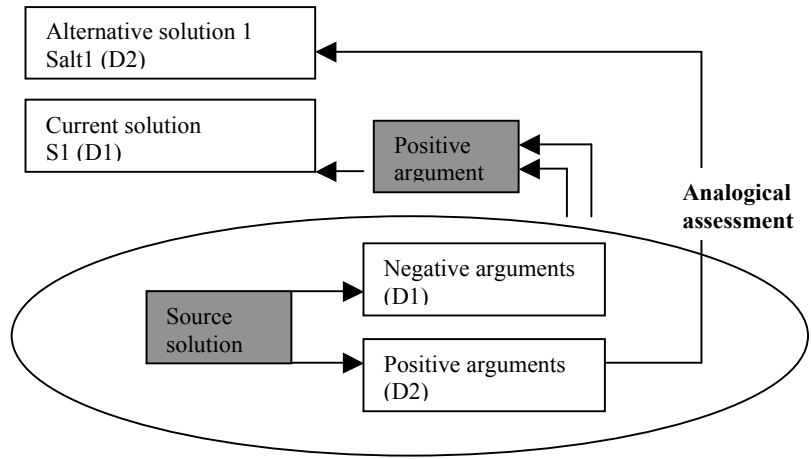

Conclusion : Verification of information on the source

*Figure 5  Comparative/analogical assessment*

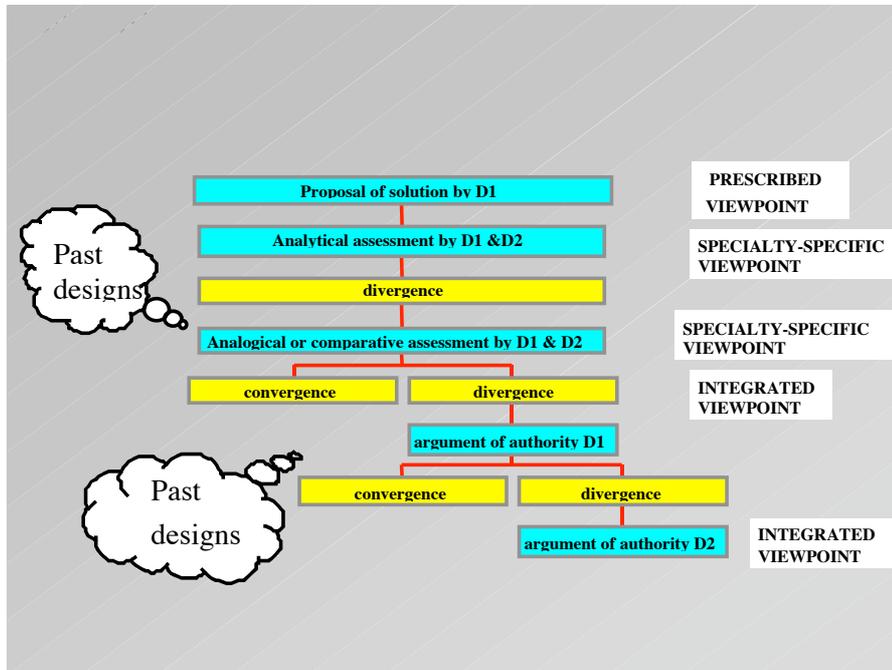

*Figure 6  Dynamics of viewpoints confrontation/integration*

**Table 1  Implicit divergence leading to constraint clarification**

| | STRUCTURE/SYSTEM-INTALLATION MEETING | | | |
|---|---|---|---|---|
| Code unit | Designer | Argument | Constraints | Diverging/ converging moves |
| 75 | SI1 | Proposal | | |
| … | … | … | | |
| 78 | St1 | Arg(-)14 | Stress implicit /structure implicit | |
| 79 | St2 | … | | Implicit divergence |
| 80 | St1 | Arg(-)15 | **Stress explicit** / structure implicit | |
| 81 | St2 | Arg(-)16 | **Stress explicit** /structure implicit | |
| 82 | SI1 | | | |
| 83 | SI1 | | | Explicit divergence |

KEY
STN: STRUCTURE-DESIGNER-N
SIN: SYSTEM-INSTALLATION-DESIGNER-N
IN BOLD: CLARIFICATION OF CONSTRAINT

**Table 2  Reinforcement of consensus by clarifying constraints**

| | STRUCTURE/SYSTEM-INSTALLATION MEETING | | |
|---|---|---|---|
| Code unit | Designer | Argument | Constraints |
| 78 | St1 | Arg(-)3 | Time explicit /cost implicit/ program-Study implicit |
| 80 | St1 | Arg(-)4 | Stress explicit |
| 81 | St2 | Arg(-)5 | Time explicit /cost implicit/ **program-Study explicit** |
| 82 | St2 | Arg(-)6 | Stress explicit / structure implicit |

KEY
STN: STRUCTURE-DESIGNER N
IN BOLD: CLARIFICATION OF CONSTRAINT

**Table 3 Contraints ranking from the most important (level 1) to least important (level 4) for two disciplines**

In italics: hydraulics-system-installation-designers specific constraints
In bold: structure-designers specific constraints

|         | *Hydraulics designers*                    | **Structure designers**   |
|---------|-------------------------------------------|---------------------------|
| Level 1 | *System installation* production time     | Maintainability           |
| Level 2 | Maintainability                           | **Structure Stress**      |
| Level 3 | Growth of problem                         | production                |
| Level 4 | *Frontier*                                | *System Installation*     |
| Level 5 | **Structure Stress**                      |                           |

**Table 4 Number of occurrence of assessment modes in meetings**

| Meetings | Design Assessment Sequence number | Number of alternative solutions | Analytical Assessment mode | Analogical /analytical Assessment mode | Analogical /comparative Assessment mode | Comparative /analytical Assessment mode | Convergence (Yes or No) |
|---|---|---|---|---|---|---|---|
| Structure /fuel SI | 1 | 5 | 3 | 0 | 1 | 0 | Yes |
| Structure /electricity SI | 2 | 0 | 1 | 1 | 0 | 0 | Yes |
|  | 3 | 1 | 1 | 0 | 1 | 0 | No |
| Structure /hydraulics SI | 4 | 4 | 5 | 0 | 2 | 0 | Yes |
|  | 5 | 4 | 2 | 0 | 0 | 2 | Yes |
|  | 6 | 4 | 2 | 1 | 0 | 1 | No |
|  | 7 | 0 | 1 | 1 | 0 |  | Yes |
| Structure /flight control SI | 8 | 15 | 15 | 2 | 0 | 1 | Yes |
| Structure /production | 9 | 5 | 4 | 4 | 0 | 0 | Yes |
| Structure /maintenance | 10 | 21 | 19 | 4 | 0 | 5 | Yes |